\documentclass[12pt]{article}

\usepackage{amsfonts}  
\newtheorem{definition}{Definition}[section]
\newtheorem{theorem}[definition]{Theorem}

\begin{document}

\begin{center}
{\LARGE  {\sffamily 
Nested Bethe ansatz for ${\cal Y}(gl({\mathfrak n}))$ open spin chains\\[1.2ex]
 with diagonal boundary conditions}}\\[4.2ex]
{\large \underline{S. Belliard} \footnote{belliard@bo.infn.it}
\\
\textsl{Istituto Nazionale di Fisica Nucleare, Sezione di Bologna, Italy}
\\[.42cm]
E. Ragoucy\footnote{eric.ragoucy@lapp.in2p3.fr}
\\
 \textsl{Laboratoire de Physique Th{\'e}orique LAPTH, UMR 
5108 
du CNRS, associ{\'e}e {\`a} l'Universit{\'e} de
Savoie, Annecy-le-Vieux Cedex, France.}
}
\end{center}

\begin{abstract}
In this proceeding 
we present the nested Bethe ansatz for open spin chains 
of XXX-type, with 
arbitrary representations (i.e. `spins') on each site of the chain 
and diagonal boundary matrices $(K^+(u),K^-(u))$. 
The nested Bethe anstaz applies for a general $K^-(u)$, but a 
particular form of the $K^+(u)$ matrix. 
We give the eigenvalues, Bethe equations and 
the form of the Bethe vectors for the corresponding models.
The Bethe vectors are expressed using a trace formula.
\end{abstract}

\section{Introduction}

Recently we proposed a unified formulation for Nested Bethe ansatz for closed and open spin chains
with "quantum group" \cite{D1,FRT,STF,Fad}, or "reflexion algebra" \cite{Ch,S} related to $gl({\mathfrak n})$ and $gl({\mathfrak n}|{\mathfrak m})$ Lie algebras \cite{BR1,BR2}.
In this proceeding we focus to open anisotropic spin chains, or of XXX-type, related to the Yangian and the reflection algebra.
More precisely we give Bethe Vectors, eigenvalues and Bethe equations
for the 'Universal' transfer matrix, an operator over the tensor product 
of $L$  highest weight representations of the Yangian. These representations
are chosen in the set of irreducible finite dimensional representations.
This approach generalises the fundamental case studies in \cite{GM} and needs
deeper analysis of the algebraic structure of the Reflection Algebra to be perform.  
The main points of this work are the explicit construction of the Bethe vectors as 
a trace formula, the construction of the Bethe vectors using embedding between different rank of reflection algebras and the proof of the validity of the Bethe Ansatz for arbitrary irreducible finite dimensional representations
 (up to some constraint
on the boundary). 
We give here a proof by increasing recursion contrary to the decreasing proof of \cite{BR2}.

The plan of the proceeding is the following. First we recall definitions and property of the
Yangian ${\cal Y}_{\mathfrak n}$ and the reflection algebra ${\mathfrak D}_{\mathfrak n}$. Then we give the finite dimensional representations of  ${\cal Y}_{\mathfrak n}$ 
and deduce the ones of ${\mathfrak D}_{\mathfrak n}$.
Next we recall the Bethe ansatz for ${\mathfrak n}=2$.
To perform the Nested Bethe anstaz, we present embedding for the reflection algebra of different rank (valid up to some quotient)
and the Bethe vectors in two forms, trace formula and recursion formula.
Then we give the proof of the Nested Bethe anstaz for open spin chains with $K^+(u)={\mathbb I}$ (the other possibility is briefly discussed).
To finish we give some open problem from this result.

\section{RTT Formalism and Yangian}

Periodic anisotropic spin chains are closely related to the Yangian ${\cal Y}_{{\mathfrak n}}$.
Among these realisations \cite{D2,D1}, the so called RTT (or  FRT)  \cite{FRT} formalism 
is the more efficient to construct the conserved quantities of the model.
These quantities belong to a abelian sub-algebra of the Yangian,
and are generated from transfer matrix.
The explicit construction of local Hamiltonian relevant for physics applications from
the transfer matrix is not easy to do in great generality, and we will focus only on its 
study.   

Let us recall the definition of the Yangian in this RTT formalism.
${\cal Y}_{{\mathfrak n}}$ is an unital associative infinite dimensional algebra generated by: 
\begin{eqnarray}
\{ t_{ij}^{(p)}\,; \, i,j=1,\dots,n\, ; \,p \in \mathbb N/\{0\}\}
\end{eqnarray}
We gather the ${\cal Y}_{{\mathfrak n}}$ generators for same $i,j$ into a
a formal series of $u^{-1}$ and then put it in an ${\mathfrak n} \times {\mathfrak n}$  matrix 
acting in an auxiliary space ${\cal V}={\mathbb C}^{{\mathfrak n}}$. We obtain the monodromy matrix:
\begin{eqnarray}
T(u)&=& \sum_{i,j=1}^{{\mathfrak n}}  E_{ij} \otimes t_{ij}(u)
\in  End({\cal V}) \otimes {\cal Y}_{\mathfrak n} \\ 
t_{ij}(u)&=&\delta_{ij}+\sum_{n=1}^\infty t_{ij}^{(n)} u^{-n}
\end{eqnarray}
where $E_{ij}$ are ${\mathfrak n} \times {\mathfrak n}$  matrices with $1$ at the intersection of line $i$ and
column $j$ and $0$ otherwise. 
The commutation between elements of ${\cal Y}_{{\mathfrak n}}$ are given by the RTT relations:
\begin{eqnarray}
  R_{12}(u-v)\ T_{1}(u)\ T_{2}(v)=T_{2}(v)\ T_{1}(u)\ R_{12}(u-v)\, \in End({\cal V}) \otimes End({\cal V}) \otimes {\cal Y}_{\mathfrak n}. 
  \label{RTT}
\end{eqnarray}
where indices $1,2$ label the auxiliary spaces where the operators act non trivially.
The matrix $R \in End({\cal V}) \otimes End({\cal V})$ 
is the rational solution of the Yang-Baxter equation:
\begin{eqnarray}
R_{12}(u_{1}-u_{2})\ R_{13}(u_{1}-u_{3})\
R_{23}(u_{2}-u_{3})=R_{23}(u_{2}-u_{3})\ R_{13}(u_{1}-u_{3})\
R_{12}(u_{1}-u_{2}),\\
R_{12}(u)=u{\mathbb I}\otimes{\mathbb I}-\hbar P_{12},\quad
P_{12}=\sum_{i,j=1}^{\mathfrak n} E_{ij}\otimes E_{ji},\quad \qquad \qquad \qquad
  \label{YBE}
\end{eqnarray}
writen in auxiliary space 
$End({\cal V}) \otimes End({\cal V}) \otimes End({\cal V})$.
This condition is equivalent to the associativity for the product of monodromy matrices.
The $R$-matrix satisfies unitarity relation,
\begin{eqnarray}
R(u) R(-u) &=&(u-\hbar)(-u-\hbar) \,{\mathbb I} \otimes {\mathbb I}\,,
\label{Unit}
\end{eqnarray}
crossing unitarity, with $t$ transposition of the first space,
\begin{eqnarray}
  R^{t}(u)R^t(-u+{\mathfrak n} \hbar) 
&=&u(-u+{\mathfrak n} \hbar) \,{\mathbb I} \otimes {\mathbb I}\,,
  \label{CrossU}
\end{eqnarray}
and is $GL({\mathfrak n},{\mathbb C})$ group invariant:
\begin{eqnarray}
[R(u),M\otimes M]=0,  \quad M \in GL({\mathfrak n},{\mathbb C})
\end{eqnarray}
The transfer matrix is defined as the trace over auxiliary space of the monodromy matrix $t(u)=tr(T(u))$ and commutes for 
different values of the formal variable $u$.
\begin{eqnarray}
[t(u),t(v)]=0
\end{eqnarray}
This is the main object to study for periodic anisotropic spin chains or more generally for two dimensional quantum integrable models related to the Yangian  \cite{Fad,KR,MTV,BR1,STF}.

The Yangian have the following automorphisms:
\\
-Shift of the spectral parameter :
\begin{eqnarray}
\sigma_a 
\,:\quad T(u)\,\to\,
T(u+a)
\end{eqnarray}
-Product by scalar function:
\begin{eqnarray}
f: T(u) \to f(u)T(u)
\end{eqnarray}

antimorphisms:
\\
- Matrix inversion:
\begin{eqnarray} \quad inv:\quad  T(u)\,\to\,T^{-1}(u)=
\sum_{i,j=1}^{{\mathfrak n}} E_{ij}\otimes t'_{ij}(u)\,.
\end{eqnarray}
- Spectral parameter inversion:
\begin{eqnarray} \quad inv:\quad  T(u)\,\to\,T(-u). \end{eqnarray}

and an Hopf algebra structure $(\Delta,S,\epsilon)$ with the coproduct define as:
\begin{eqnarray}
\Delta \quad: \quad \Delta(T(u)) =
T(u)\dot{\otimes}T(u) = \sum_{i,j,k=1}^{{\mathfrak n}} \,
E_{ij}\otimes t_{ik}(u)\otimes t_{kj}(u).\end{eqnarray}
More generally, one defines recursively for $L\geq 2$,
\begin{eqnarray}
\Delta^{(L+1)}=(\mbox{id}^{\otimes (L-1)}\otimes \Delta)\circ\Delta^{(L)}\
:\ {{\cal Y}_{{\mathfrak n}}} \to {{\cal Y}_{{\mathfrak n}}}^{\otimes (L+1)},
\end{eqnarray}
with $\Delta^{(2)}=\Delta$ and $\Delta^{(1)}=\mbox{id}$. The map 
$\Delta^{(L)}$ is an algebra homomorphism.

The Yangian has the universal enveloping algebra ${\cal U}(gl({\mathfrak n}))$ as an Hopf subalgebra, the embedding
is given by ${\cal E}_{ji} \to t^{(1)}_{ij}$.
Where ${\cal E}_{ij}$ are the generators of ${\cal U}(gl({\mathfrak n}))$ with commutation relations:
\begin{eqnarray}
[{\cal E}_{ij},{\cal E}_{kl}]=\delta_{jk}{\cal E}_{ik}-\delta_{il}{\cal E}_{kj}
\end{eqnarray}
The evaluation homomorphism $ ev: {\cal Y}_{\mathfrak n} \to {\cal U}(gl({\mathfrak n}))$ is given by:
\begin{eqnarray}
 ev &:& t^{(1)}_{ij} \to {\cal E}_{ji}\nonumber \\
 ev &:& t^{(p)}_{ij} \to 0 , \quad p>1
\end{eqnarray}
This evaluation homomorphism is the key ingredient to construct finite dimensional
representation of $ {\cal Y}_{\mathfrak n}$ \cite{D2,M1,CP1}.  

Let us introduce some notation used for $R$ matrices in this paper:
\\
 -The 'normalized' $R$-matrices: 
\begin{eqnarray}
\mbox{${\mathbb R}$}(u)=\frac{R(u)}{(u-\hbar)}\,
\mbox{with} 
\mbox{${\mathbb R}$}(u)\mbox{${\mathbb R}$}(-u)={\mathbb I}\otimes{\mathbb I}.
\label{Runit}
\end{eqnarray}
-And the `reduced' $R$-matrices $R^{(k,p)}(u)$:
\begin{eqnarray}
R^{(k,p)}(u)&=&
\left({\mathbb I}^{(k)} \otimes {\mathbb I}^{(p)}\right) R(u)
\left({\mathbb I}^{(k)} \otimes {\mathbb I}^{(p)}\right), \quad
\mbox{with} \quad{\mathbb I}^{(k)}=\sum_{i=k}^{{\mathfrak n}}E_{ii}\,,\nonumber \\
R^{(k)}(u)&=&R^{(k,k)}(u).
\label{eq:Ik}
\end{eqnarray}
We have $R^{(1)}(u)=R(u)$, and more generally 
$R^{(k)}(u)$ 
corresponds to the $R$-matrix of
${\cal Y}_{{\mathfrak n}-k+1}$.

\section{Reflection algebra and $K(u)$ matrices} 
The  ${\cal Y}_{{\mathfrak n}}$ algebra is enough to 
construct a transfer matrix leading to
periodic models, but
in the context of open spin chains, one needs another algebra, 
the reflection algebra
${\mathfrak D}_{{\mathfrak n}}$ \cite{Ch}, which turns out to be a subalgebra of ${\cal Y}_{{\mathfrak n}}$. 
Indeed, physically, one can interpret the 
 RTT relation as encoding the interaction between the spins of the 
chain. Hence, it is the only relation needed to describe a periodic 
chain. On the other hand, in the case of open chain, the interaction with the 
boundaries has to be taken into account. 
Following the seminal paper \cite{S}, 
one constructs the reflection algebra and the 
dual reflection equation
 for the boundary scalar matrices $K^-(u)$ and $K^+(u)$. 
We first define the matrix $K^-(u)$ to be the solution of the 
reflection equation in $End({\cal V})\otimes End({\cal V})$:
\begin{eqnarray}
  R_{12}(u_{1}-u_{2}) K^-_{1}(u_{1})
  R_{12}(u_{1}+u_{2}) K^-_{2}(u_{2})=
  K^-_{2}(u_{2})R_{12}(u_{1}+u_{2})
  K^-_{1}(u_{1})R_{12}(u_{1}-u_{2}) .
  \label{ERK}
\end{eqnarray}
The scalar solutions to the 
reflection equation have been classified using the $GL({\mathfrak n},{\mathbb C})$ invariance of $R$ matrix \cite{AACDFR}.
The diagonal solutions take the form (up to normalisation),
\begin{eqnarray}
K^-(u)=diag(\underbrace{u-c_-,\dots,u-c_-}_a,
\underbrace{-u-c_-,\dots,-u-c_-}_{{\mathfrak n}-a}) =\sum_{i=1}^{\mathfrak n} \kappa^-_i(u)E_{ii}
\label{eq:Kdiag}
\end{eqnarray}
where $c_{-}$ is a free complex parameter and $a$ is an integer.
From this $K^-(u)$ matrix and the monodromy matrix $T(u)$ of ${\cal Y}_{\mathfrak n}$,
we can construct the monodromy matrix of ${\mathfrak D}_{\mathfrak n} \subset {\cal Y}_{\mathfrak n}$:
\begin{eqnarray}
D(u)&=&T(u)\,K^-(u)\,T^{-1}(-u) 
\ =\ \sum_{i,j=1}^{{\mathfrak n}} d_{ij}(u)\otimes E_{ij}, 
\label{D-from-T}\\
d_{ij}(u) &=& \sum_{a=1}^{{\mathfrak n}}\,
\kappa^-_a(u)t_{ia}(u)t'_{aj}(-u).
\end{eqnarray}
From (\ref{RTT})  and  (\ref{ERK}), one can prove that $D(u)$ also
satisfies the reflection equation in $End({\cal V})\otimes End({\cal V})\otimes {\mathfrak D}_{\mathfrak n}$: 
\begin{eqnarray}
R_{12}(u_{1}-u_{2}) D_{1}(u_{1})
R_{12}(u_{1}-u_{2}) D_{2}(u_{2})=
D_{2}(u_{2})R_{12}(u_{1}-u_{2})
D_{1}(u_{1})R_{12}(u_{1}-u_{2}). 
\label{ER}
\end{eqnarray}
The algebra ${\mathfrak D}_{{\mathfrak n}}$ is a left coideal \cite{MoRa} of the 
algebra ${\cal Y}_{{\mathfrak n}}$ with coproduct action:
\begin{eqnarray}
\Delta (D_{[2]}(u))=T_{[1]}(u)D_{[2]}(u)T_{[1]}^{-1}(-u) \in End({\cal V})\otimes{\cal Y}_{{\mathfrak n}}\otimes{\mathfrak D}_{{\mathfrak n}} 
\label{eq:deltaD}
\end{eqnarray}
where $[1]$ and $[2]$ labels the algebras ${\cal Y}_{{\mathfrak n}}$ and ${\mathfrak D}_{{\mathfrak n}}$, respectively.

To construct commuting transfer matrices we introduced a dual equation in $End({\cal V})\otimes End({\cal V})$ :
\begin{eqnarray}
&&R_{12}(u_{2}-u_{1}) (K^+_{1}(u_{1}))^{t_1}
R_{12}(-u_{1}-u_{2}+{\mathfrak n} \hbar)
(K^+_{2}(u_{2}))^{t_2}=\nonumber \\
&& 
 (K^+_{2}(u_{2}))^{t_2}R_{12}(-u_{1}-u_{2}+{\mathfrak n} \hbar)
  (K^+_{1}(u_{1}))^{t_1}R_{12}(u_{2}-u_{1}). 
  \label{ERD}
\end{eqnarray}
From isomorphism of the refection equation and dual reflection equation, one can construct solutions to the dual reflection equation from
$K^-(u)$:
\begin{eqnarray}
(K^+(u))^t=K^-(-u+\frac{{\mathfrak n}}{2}\hbar)\,,
\end{eqnarray}
With $D(u)$ and $K^+(u)$ one constructs the transfer matrix: 
\begin{eqnarray}
d(u) = tr(K^+(u) D(u)). 
\end{eqnarray}
The reflection equation and its dual form ensure the commutation 
relation:
\begin{eqnarray}
[d(u),d(v)] = 0. 
\end{eqnarray}
Thus, $d(u)$ generates (via an expansion in $u^-1$) a set of commuting conserved quantities and is related to boundaries anisotropy spin chains models and more generally to  boundaries quantum integrable models related to ${\mathfrak D}_{\mathfrak n}$.

\section{Highest weight representations\label{sec:hw}}

The fundamental point in using the ABA is to know a pseudo-vacuum 
for the model. In the mathematical framework 
it is equivalent to know a highest weight 
 representation for the algebra which underlies the model. 
Since the generators of the algebra ${\mathfrak D}_{{\mathfrak n}}$ can be constructed 
from the ${\cal Y}_{{\mathfrak n}}$ ones, see eq. (\ref{D-from-T}), 
we first describe how to construct highest repesentations for the infinite 
dimensional algebras
${\cal Y}_{{\mathfrak n}}$ from highest  weight representations of the 
finite dimensional Lie algebras $gl({\mathfrak n})$.
Next, we show how these representations induce (for diagonal $K^-(u)$  
matrix) a representation for ${\mathfrak D}_{{\mathfrak n}}$ with same
highest weight vector.

\begin{definition}
A representation  of ${\cal Y}_{{\mathfrak n}}$ is called \textit{highest 
weight} 
if there exists a nonzero vector $\Omega$ 
such that,
\begin{eqnarray}
t_{ii} (u)\, \Omega = \lambda_{i}(u)\, \Omega \quad \mbox{and} \quad
t_{ij} (u)\, \Omega = 0 \ \mbox{ for }\ i>j,   
\label{higvectU}
\end{eqnarray}  
for some scalars $\lambda_{i}(u)$ $\in$ ${\mathbb C}$ $[[u^{-1}]]$.
$\lambda(u)= (\lambda_{1}(u),\dots,\lambda_{{\mathfrak n}}(u))$ is 
called 
the highest weight and $\Omega$ the highest weight vector.
\end{definition}
It is known that
any finite-dimensional irreducible representation of 
${\cal Y}_{{\mathfrak n}}$ is  highest weight and that it contains a unique 
 (up to scalar multiples) highest weight vector \cite{CP1,M1}.

To construct such representations, one uses the evaluation morphism, 
which relates the infinite dimensional algebra ${\cal Y}_{{\mathfrak n}}$ to its
finite dimensional subalgebra ${\cal U}(gl({\mathfrak n}))$ and a finite dimensional irreducible highest weight 
representation
$\pi_{\mu}: {\cal U}(gl({\mathfrak n})) \to End({\cal V}_\mu)$ with highest weight $\Omega \in {\cal V}_\mu$: 
\begin{eqnarray}
\pi_{\mu}({\cal E}_{ij})\Omega=0, \quad 1\leq i<j \leq n , \quad
\quad \pi_{\mu}({\cal E}_{ii})\Omega=\mu_i\Omega, \quad 1\leq i \leq n,\quad
\mu_i-\mu_{i+1} \in \mathbb Z_+
\end{eqnarray}

The evaluation representations of ${\cal Y}_{{\mathfrak n}}$  are constructed by the following composition
of maps: 
\begin{eqnarray}
\rho^\mu_{a}=\pi_{\mu} \circ\, ev \circ \sigma_a :\quad  {\cal Y}_{{\mathfrak n}}\stackrel{ \sigma_a}{\longrightarrow} {\cal Y}_{{\mathfrak n}}\ 
\stackrel{ev}{\longrightarrow}\ {\cal U}(gl({\mathfrak n}))\  
\stackrel{\pi_{\mu}}{\longrightarrow}\ End({\cal V}_{\lambda})\,.  
\end{eqnarray}
The weight of this evaluation representation is given by
$\lambda(u)=\big(\lambda_{1}(u),\ldots,\lambda_{{\mathfrak n}}(u)\big)$, 
with:
\begin{eqnarray}
\lambda_{j}(u) =
 u-a-\,\hbar\,\mu_j \quad ,\quad j=1,\ldots,{\mathfrak n},
\label{eq:Lambda-eval}
\end{eqnarray}

More generally, one constructs tensor products of evaluation 
representations using the coproduct of ${\cal Y}_{{\mathfrak n}}$,
\begin{eqnarray}
\Big(\otimes_{i=1}^{L}\,\rho^{\mu^{\langle 
i\rangle}}_{a_i}\Big)\, \circ 
\Delta^{(L)}\Big(T(u)\Big) = \rho^{\mu^{\langle 
1\rangle}}_{a_{1}}\Big(T(u)\Big)\dot{\otimes}\,
\rho^{\mu^{\langle 2\rangle}}_{a_{2}}\Big(T(u)\Big) 
\dot{\otimes}\cdots 
\dot{\otimes}\rho^{\mu^{\langle L\rangle}}_{a_{L}}\Big(T(u)\Big),
\label{eq:mono-repr}
\end{eqnarray}
where $\mu^{\langle i\rangle}=(\mu^{\langle 
i\rangle}_{1},\ldots,
\mu^{\langle i\rangle}_{{\mathfrak n}})$, 
$i=1,\ldots,L$, are the weights of the ${\cal U}(gl({\mathfrak n}))$ 
representations. 
This provides a ${\cal Y}_{{\mathfrak n}}$ representation with weight,
\begin{eqnarray}
\lambda_j(u)=\prod_{i=1}^{L}\lambda^{\langle i\rangle}_j(u)
\,,\qquad j=1,\ldots,{\mathfrak n},
\label{VP}
\end{eqnarray}
where $\lambda^{\langle i\rangle}_j(u)$ have the form 
(\ref{eq:Lambda-eval}).
Evaluation representations are central in the study of representations
 because all finite dimensional irreducible representations of ${\cal Y}_{{\mathfrak n}}$ 
can be constructed from tensor products of evaluation 
representations (see \cite{BR1} for references).


To obtain representation of $D(u)$ we also need to give
$T^{-1}(u)$ in term of the $T(u)$ elements.
It could be done
using the quantum determinant
$qdet(T(u))$ and the comatrix
$\widehat T(u)$ see \cite{M2}.

The  quantum determinant 
$qdet(T(u))$ which generates the center of ${\cal Y}_{\mathfrak n}$ is defined as:
\begin{eqnarray}
qdet(T(u))=\sum_{\sigma \in S_{\mathfrak n}}sign(\sigma)\,
\prod_{i=1}^{{\mathfrak n}}t_{i \sigma(i)}(u+(i-{\mathfrak n})\hbar), 
\end{eqnarray}
where $S_{\mathfrak n}$ is the permutation group of ${\mathfrak n}$ elements and $\sigma$
a permutation with signature $sgn(\sigma)$.

The quantum comatrix $\widehat T(u)$
satisfies:
\begin{eqnarray}
\widehat T(u)\,T(u-({\mathfrak n}-1)\hbar)=qdet(T(u)),
\end{eqnarray} 
this equation allows to relate $T^{-1}(u)$ to 
$\widehat T(u)$: 
\begin{eqnarray}
T^{-1}(u)=\frac{\widehat T(u+({\mathfrak n}-1)\hbar)}{qdet(T(u+({\mathfrak n}-1)\hbar))}.
\end{eqnarray}

From the exact form of $\widehat T(u)$ in term of $t_{ij}(u)$ we can find that $\Omega$ is also
a highest weight vector for  $T^{-1}(u)$ with weights :
\begin{eqnarray}
t'_{ii}(u)\,\Omega&=&\lambda'_i(u)\,\Omega, \quad
\lambda'_i(u)=\left( 
\prod_{k=1}^{i-1}\frac{\lambda_k(u+k\hbar)}{\lambda_k(u+(k-1)\hbar)} 
\right)\frac{1}{\lambda_i(u+(i-1)\hbar)}.
\end{eqnarray}

The representations of the reflection algebra ${\mathfrak D}_{\mathfrak n}$, could be study from previous results \cite{MoRa}.
For $K^-(u)$ diagonal, the finite dimensional irreducible highest weight representations follow from the ones of ${\cal Y}_{\mathfrak n}$
and lead to the following theorem:
  
\begin{theorem}\label{theo:valB}
If $\Omega$ is a highest weight vector of ${\cal Y}_{{\mathfrak n}}$, with 
eigenvalue
$( \lambda_{1}(u),\ldots, \lambda_{{\mathfrak n}}(u))$, then, when $K(u)$ 
is a diagonal
matrix with $\kappa_i(u)$ diagonal elements, $\Omega$ is also a highest weight 
vector for ${\mathfrak D}_{{\mathfrak n}}$,
\begin{eqnarray}
d_{ij}(u)\,\Omega =0 \quad \mbox{for} \quad i>j,
\quad \mbox{and} \quad
d_{ii}(u)\,\Omega =\Lambda_i(u,\{\kappa^-(u),\lambda(u),\lambda'(-u)\})\,\Omega, 
\end{eqnarray}
with eigenvalues:
\begin{eqnarray}
\Lambda_i(u)&=&{\cal K}_{i}(u)\,\lambda_{i}(u)\,\lambda'_{i}(-u)
-\sum_{k=1}^{i-1}
\frac{\hbar}{2u-\frac{(k-1)\hbar}{2}}\,
{\cal K}_{k}(u)\,\lambda_{k}(u)\,\lambda'_{k}(-u),
\label{eq:valprD}\\
{\cal K}_{i}(u)&=&\kappa_{i}(u)+\sum_{k=1}^{i-1}\kappa_{k}(u)
\frac{\hbar}
{2u-\frac{(i-2)\hbar}{2}}
\label{eq:grKapp}
\end{eqnarray}
\end{theorem}

Now we can introduce what we call 'general transfer matrix' ${\mathfrak d}(u; L,\{a\},\{\mu\})$:
\begin{eqnarray}
{\mathfrak d}(u; L,\{a\},\{\mu\})=\Big(\otimes_{i=1}^{L}\,\rho^{\mu^{\langle 
i\rangle}}_{a_i}\Big)\, \circ 
\Delta^{(L)}(d(u))
\end{eqnarray}
The next section we will give the proof of the Nested Bethe ansatz for this 'general transfer matrix'.
To simplify notation we will use $d(u)$ for ${\mathfrak d}(u; L,\{a\},\{\mu\})$ in the next sections. 

\section{Algebraic Bethe ansatz for ${\mathfrak D}_{{\mathfrak n}}$ with ${\mathfrak n}=2$ 
\label{sec:ABA}}
In this section, we remind the framework of the Algebraic Bethe 
Ansatz (ABA) \cite{STF}
introduced  in order to compute transfer matrix
eigenvalues and eigenvectors.  
The method follows the same steps as the closed chain case, up to a 
preliminary step. 
We will only consider the case $K^+(u)={\mathbb I}$ which is relevant for the Nested Bethe anstaz. 
In the open case the transfer matrix hs the form: 
\begin{eqnarray}
d(u) &=& tr(D_a(u))=d_{11}(u)  + d_{22}(u).
\end{eqnarray}
We perform a change of basis and a shift, 
\begin{eqnarray}
d_{11}(u+\frac{\hbar}{2}) &=& \widehat d_{11}(u)\mbox{,} \quad d_{12}(u+\frac{\hbar}{2})=\widehat 
d_{12}(u)
\mbox{,} \quad 
d_{21}(u+\frac{\hbar}{2})=\widehat d_{21}(u),\\
d_{22}(u+\frac{\hbar}{2}) &=& \widehat d_{22}(u)-\frac{\hbar}{2u}\,\widehat d_{11}(u).
\end{eqnarray}
This change of basis leads to symmetric exchange relations:
\begin{eqnarray}
[ \widehat d_{12}(u),\widehat d_{12}(v)]&=& 0 ,
\label{eq:d12d12hat}\\
\widehat d_{11}(u)\, \widehat d_{12}(v)&=&
\frac{(u-v+\hbar)(u+v+\hbar)}{(u-v)(u+v)}\widehat d_{12}(v)\, \widehat d_{11}(u)
-\frac{\hbar(2v+\hbar)}{2v(u-v)}\widehat d_{12}(u)\, \widehat d_{11}(v)
\nonumber \\
&& +\frac{\hbar}{u+v}\,\widehat d_{12}(u)\, \widehat d_{22}(v),
\label{ExchABF} 
\\
\widehat d_{22}(u)\, \widehat d_{12}(v)&=& \frac{(u-v-\hbar)(u+v-\hbar)}{(u-v)(u+v)}\,
 \widehat d_{12}(v) \, \widehat d_{22}(u)
+\frac{\hbar(2u-\hbar)}{2u(u-v)}\widehat d_{12}(u)\, \widehat d_{22}(v) 
\nonumber \\
 && -\frac{\hbar(2u-\hbar)(2v+\hbar)}{4uv(u+v)}\, \widehat d_{12}(u)\, \widehat d_{11}(v).
\label{ExchABFbis} 
\end{eqnarray}
In the new basis, $\Omega$ is still a pseudo-vacuum:
\begin{eqnarray}
\widehat d_{11}(u) \, \Omega  &=&\widehat{\Lambda}_1(u)\,\Omega= {\cal K}_1(u+\frac{\hbar}{2})
\lambda_1(u+\frac{\hbar}{2})\lambda'_1(-u-\frac{\hbar}{2}) \,\Omega \mbox{,} \quad
\widehat d_{21}(u) \, \Omega  \,=\,  0,\\
\widehat d_{22}(u) \, \Omega  &=& \widehat \Lambda_2(u) 
\,\Omega \,=\,\Big(\Lambda_{2}(u+\frac{\hbar}{2})+\frac{\hbar}{2u}\,
\Lambda_1(u+\frac{\hbar}{2}) \Big)\,\Omega\nonumber \\
&=& {\cal K}_2(u+\frac{\hbar}{2})
\lambda_2(u+\frac{\hbar}{2})\lambda'_2(-u-\frac{\hbar}{2}) \,\Omega 
 \,.
\end{eqnarray}
and we can use the algebraic Bethe ansatz as in the closed chain 
case. 
The transfer matrix rewrites:
\begin{eqnarray}
d(u+\frac{\hbar}{2}) =
\frac{2u-\hbar}{2u}\,
\widehat d_{11}(u)+
\widehat d_{22}(u)
\equiv \widehat d(u)
\end{eqnarray}
Applying $M$ creation operators $\widehat d_{12}(u_{j})$ on the pseudo vacuum we 
generate a Bethe vector:
\begin{eqnarray}
\Phi(\{u\}) \, = \widehat d_{12} (u_1) \dots  \widehat d_{12} (u_M)\Omega.
\end{eqnarray}
where $\{u\}=\{u_1,\dots,u_M\}$.
Demanding $\Phi(\{u\})$ to be an eigenvector of $\widehat d(u)$ leads
to a set of algebraic relations on the parameters $\{u\}$, the so-called Bethe equations:
\begin{eqnarray}
\frac{ {\cal K}_1(u_k+\frac{\hbar}{2})
\lambda_1(u_k+\frac{\hbar}{2})\lambda'_1(-u_k-\frac{\hbar}{2})}{ {\cal K}_2(u_k+\frac{\hbar}{2})
\lambda_2(u_k+\frac{\hbar}{2})\lambda'_2(-u_k-\frac{\hbar}{2})} &=&
\frac{2u_k}{2u_k+\hbar}\prod_{i\neq k}^{l}\frac{(u_k-u_i-\hbar)(u_k+u_i-\hbar)}{(u_k-u_i+\hbar)(u_k+u_i+\hbar)},
\end{eqnarray}
Then, the eigenvalues of the transfer 
matrix read:  
\begin{eqnarray}
d(u)\,\Phi(\{u\}) &=& \Lambda(u)\, \Phi(\{u\}), \\
\Lambda(u) &=& 
\frac{2u-2h}{2u-\hbar}\,
{\cal K}_1(u)
\lambda_1(u)\lambda'_1(-u)\prod_{k=1}^{M} 
 \frac{(u-u_k+\frac{\hbar}{2})(u+u_k+\frac{\hbar}{2})}{(u-u_k-\frac{\hbar}{2})(u+u_k-\frac{\hbar}{2})}\nonumber \\
&&+ {\cal K}_2(u)
\lambda_2(u)\lambda'_2(-u)\,
 \prod_{k=1}^{M} \frac{(u-u_k-\frac{3\hbar}{2})(u+u_k-\frac{3\hbar}{2})}{(u-u_k-\frac{\hbar}{2})(u+u_k-\frac{\hbar}{2})}\,.
\end{eqnarray}
Note that Bethe equations correspond to the vanishing
of the residue of  $\Lambda(u)$ at $u=u_j+\frac{\hbar}{2}$. 

\section{Nested Bethe ansatz\label{sec:NBA}}
In this section we will give the step for a direct recursion for the Bethe equations and eigenvalues
of a "general open spin chain" of rank ${\mathfrak n}+1$.
This proof uses the knowledge of the recursion formula for the Bethe vectors and the embedding
$ {\mathfrak D}_{{\mathfrak n}} \to  {\mathfrak D}_{{\mathfrak n}+1}$. More precisely, one has to consider the quotient of ${\mathfrak D}_{{\mathfrak n}+1}$ 
by ${\cal I}$, the left ideal generate by $\{d_{ij}(u), i>j\}$ .
First we give the theorem about this embedding and next we prove the Nested Bethe ansatz 
for  ${\mathfrak D}_{{\mathfrak n}+1}$  from the ${\mathfrak D}_{{\mathfrak n}}$ one.
This formulation gives an alternative proof of the one given in \cite{BR2} for a more general case.

\subsection{Embeddings of ${\mathfrak D}_{{\mathfrak n}}$ algebras}

The algebraic cornerstone for the nested Bethe ansatz is a 
recursion relation on
 the ${\mathfrak D}_{{\mathfrak n}}$ algebraic structure:
 \begin{eqnarray}
 {\mathfrak D}_{{\mathfrak n}} \to {\mathfrak D}_{{\mathfrak n}-1} \to \dots \to {\mathfrak D}_{3} \to {\mathfrak D}_{2}
 \end{eqnarray}
In this section we give two important 
properties of the algebra
${\mathfrak D}_{{\mathfrak n}}$, described in the following theorem:

\begin{theorem}
\label{theo:embedding}
For $k=1,2,\ldots,{\mathfrak n}-1$, let $F^{(k)}$ be a linear combination of 
 $d_{i_1j_1}(u_1)\dots d_{i_lj_l}(u_l)$
with all indices $k-1<i_{p}\leq j_{p}$, and let
${\cal I}$ be the left ideal generated by 
$d_{ij}(u)$ for $i>j$.
Then,
we have the following properties:
\begin{eqnarray}
 d_{ij}(u)\,F^{(k)}&\equiv&0 \quad\mbox{mod ${\cal I}$}\,,\quad \mbox{for} \quad i>j \quad \mbox{and} \quad
j<k,\\
\null [d_{ii}(u)\,,F^{(k)}] &\equiv&0 \quad\mbox{mod ${\cal I}$}\,, \quad\mbox{for}\quad i<k.
\label{com-LC}
\end{eqnarray}
and the generators:
\begin{eqnarray}
\widehat D^{(k)}(u)&=&
\sum_{i,j=k}^{{\mathfrak n}} E_{ij} \otimes d^{(k)}_{ij}(u), \\
d^{(k)}_{ij}(u)&=&d_{ij}(u+\frac{( k - 1)\hbar}{2})+\delta_{ij}\sum_{a=1}^{k-1} \frac{\hbar}{2u} \,
d_{aa}(u+\frac{( k - 1)\hbar}{2})\,
\label{TF-D}
\end{eqnarray}
satisfy in ${\mathfrak D}_{{\mathfrak n}}/{\cal I}$ the reflection equation for 
${\mathfrak D}_{{\mathfrak n}-k+1}$:
\begin{eqnarray}
R^{(k)}_{12}(u_{1}-u_{2}) \widehat D^{(k)}_{1}(u_{1})
R^{(k)}_{12}(u_{1}+u_{2})\widehat D^{(k)}_{2}(u_{2})
\equiv
\widehat D^{(k)}_{2}(u_{2}) R^{(k)}_{12}(u_{1}+u_{2})
\widehat D^{(k)}_{1}(u_{1}) R^{(k)}_{12}(u_{1}-u_{2})  \nonumber
\end{eqnarray}
\end{theorem}

Let us give two useful relations from this theorem for the Nested Bethe Ansatz:

-The action of $d^{(k)}_{kk}(u)$ on $\Omega$ :
\begin{eqnarray}
 d^{(k)}_{kk}(u)\Omega={\cal K}_{k}(u+\frac{(k-1)\hbar}{2})\lambda_k(u+\frac{(k-1)\hbar}{2})\lambda'_k(-u-\frac{(k-1)\hbar}{2})
 \end{eqnarray}

-The embedding $\tau: {\mathfrak D}_{{\mathfrak n}}/{\cal I}_{\mathfrak n} \to  {\mathfrak D}_{{\mathfrak n}+1}/{\cal I}_{{\mathfrak n}+1}$ given by:
\begin{eqnarray}
\tau(d_{ij}(u))=d^{(2)}_{i+1j+1}(u)=d_{i+1j+1}(u+\frac{\hbar}{2})+\delta_{i,j}\frac{\hbar}{2u}d_{11}(u+\frac{\hbar}{2})
\end{eqnarray} 
it follows:
\begin{eqnarray}
\tau(d^{(k)}_{ij}(u))=d^{(k+1)}_{i+1,j+1}(u)  \label{tau-bis}
\end{eqnarray}
This morphism will be crucial for the computation of the Nested Bethe ansatz. We will use it in the form:
 \begin{eqnarray}
 \tau(D(u))=\widehat D^{(2)}(u) \label{tau}
 \end{eqnarray}
Choosing the form (\ref{D-from-T}) for the operator $D$
we can compute the action of the coproduct of ${\cal Y}_{{\mathfrak n}}$ on 
$\widehat D^{(k)}(u)$.
\begin{theorem}
\label{embedding-co}
In the coset ${\cal Y}_{{\mathfrak n}}/{\cal J} \otimes {\mathfrak D}_{{\mathfrak n}}/{\cal I}$, where ${\cal J}$ is the left ideal generated by
 \begin{eqnarray}
 \{t_{ij}(u),t'_{ij}(-u), i>j\}, \nonumber
 \end{eqnarray}
 the coproduct takes the form
\begin{eqnarray}
\Delta(\widehat D_{[2]}^{(k)}(u)) &\equiv&
T_{[1]}^{(k)}(u)\widehat D_{[2]}^{(k)}(u)(T_{[1]}^{-1})^{(k)}(-u)
\mbox{mod ${\cal J}$},\\
 T^{(k)}(u)&=&
\sum_{i,j=k}^{{\mathfrak n}} E_{ij} \otimes t^{(k)}_{ij}(u) \mbox{and}
(T^{-1})^{(k)}(-u) = \sum_{i,j=k}^{{\mathfrak n}} E_{ij} \otimes 
t'^{(k)}_{ij}(-u),
\\
t^{(k)}_{ij}(u)&=&t_{ij}(u+\frac{( k - 1)\hbar}{2}) \mbox{ and}
t'^{(k)}_{ij}(-u) = t'_{ij}(-u-\frac{( k - 1)\hbar}{2}),
\end{eqnarray}
where $[1]$ labels the space ${\mathfrak D}_{{\mathfrak n}}/{\cal I}$, $[2]$ labels the 
space ${\cal Y}_{{\mathfrak n}}/{\cal J}$ and $\Delta$ is the coproduct of 
${\cal Y}_{{\mathfrak n}}$.
\end{theorem}

From this result and using the fundamental representation $\bar{\pi}_a$ of ${\cal Y}_i$ we can obtain the 
convenient relation for $i<k$ (see \cite{BR2}):
\begin{eqnarray}
\bar{\pi}^{(i)}_a(T^{(k)}(u))=\mbox{${\mathbb R}$}^{(i,k)}(u+a+\frac{(k-i)\hbar}{2})  \nonumber \\
\bar{\pi}^{(i)}_a((T^{-1})^{(k)}(-u))=\mbox{${\mathbb R}$}^{(i,k)}(u-a+\frac{(k-i)\hbar}{2}) \label{repf} 
\end{eqnarray}
This formulas will be use to prove the recursion of the Nested Bethe ansatz.

\subsection{Bethe vectors}
We present here a generalization to open spin chains of the 
recursion and trace formulas for Bethe vectors, obtained in \cite{MTV,TV2} for closed spin chains.

Let us introduce the following trace formula for the Bethe vectors (or weight function) of ${\mathfrak D}_{\mathfrak n}$ universal "diagonal" open spin chains.
 We introduce 
a family of Bethe 
parameters 
$u_{kj}$, $j=1,\ldots,M_{k}$, 
the number $M_{k}$ of these parameters being a free integer. The 
partial unions of these families will be noted as,
\begin{eqnarray}
\{u_{\ell}\}=\bigcup_{i=1}^{\ell}\,
\{u_{ij}\,,\ j=1,\ldots,M_{i}\},
\end{eqnarray}
so that the whole family of Bethe parameters is
$\{u\}=\{u_{{\mathfrak n}-1}\}$ with cardinal 
$M=\sum_{k=1}^{{\mathfrak n}-1}M_k$.
\begin{theorem} \label{trace}
We denote $A_1; \dots; A_{{\mathfrak n}-1} $ the ordered sequence of 
auxiliary spaces
$a_1^1, \dots, a_{M_1}^1$; $a_1^2, \dots, a_{M_2}^2$
$; \dots ;$ $a_1^{{\mathfrak n}-1}, \dots ,
a_{M_{{\mathfrak n}-1}}^{{\mathfrak n}-1}$. Then:
\begin{eqnarray}
\Phi^{{\mathfrak n}}_M(\{u\})\Omega = 
tr_{A_1 \ldots A_{{\mathfrak n}-1} }
\,\left(\prod_{i=1}^{{\mathfrak n}-1}
\widehat {\mathbb D}^{(i)}_{A_i}(\{u_i\})
 E_{{\mathfrak n},{\mathfrak n}-1}^{\otimes M_{{\mathfrak n}-1}} 
\otimes \dots \otimes  
E_{21}^{\otimes M_1}\right) \,\Omega, \quad
\label{eq:Phi-str}
\end{eqnarray}
where
\begin{eqnarray}
\widehat {\mathbb D}^{(i)}_{A_i}(\{u_i\})&=&\prod_{j=1}^{M_i}
\overline {\cal R}^{(i)}_{A_{<i},a^i_j}(\{u_{i-1}\},u_{ij})
\widehat{D}^{(i)}_{a^i_j}(u_{ij}+\frac{\hbar}{2})
{\cal R}^{(i)}_{a^i_j,A_{<i}}(\{u_{i-1}\},u_{ij}),
\\
\overline {\cal R}^{(i)}_{A_{<i},a^i_j}(\{u_{i-1}\},u_{ij})
 &=&\prod_{b<i}^{\longleftarrow} 
\prod_{c=1...M_b}^{\longrightarrow}
 \mbox{${\mathbb R}$}^{(i,b+1)}_{a_j^ia_c^b }\Big(u_{ij}+u_{bc}+\frac{(i-b+1)\hbar}{2}\Big),
 \\
 {\cal R}^{(i)}_{a^i_j,A_{<i}}(\{u_{i-1}\},u_{ij})
 &=&\prod_{b<i}^{\longrightarrow}
\prod_{c=1...M_b}^{\longleftarrow}
 \mbox{${\mathbb R}$}^{(i,b+1)}_{a_j^ia_c^b}\Big(u_{ij}-u_{bc}+\frac{(i-b+1)\hbar}{2}\Big),
\label{eq:bigR} 
\\
\prod_{i=1,..,n}^{\longrightarrow}X_i=X_1...X_n, && \prod_{i=1,..,n}^{\longleftarrow}X_i=X_n...X_1
\end{eqnarray} 
This formula is invariant under the same permutation of elements of $A_i$ and $\{u_{i1},\dots,u_{iM_i}\}$. 
\end{theorem}
The proof of the last assertion does not clearly appear in \cite{BR2} and will be published elsewhere.  

{From} the trace formula, we can extract a recurrent form 
for the Bethe vectors,
\begin{eqnarray}
\Phi^{{\mathfrak n}}_M(\{u\}) \Omega&=& \widehat B^{(1)}_{a^1_1}(u_{11}) \cdots 
 \widehat  B^{(1)}_{a^1_{M_1}}(u_{1M_1})\,\widehat\Psi^{(1)}_{\{u_1\}}
\Big(\Phi^{{{\mathfrak n}}-1}_{M-M_{1}}(\{u\}/\{u_1\})\Big)\Omega,
\label{BVR}\\
\widehat\Psi^{(1)}_{\{u_1\}} &=& 
v^{(2)}\,\circ \,
(\bar{\pi}^{(2)}_{u_{11}} \otimes \dots \otimes 
\bar{\pi}^{(2)}_{u_{1M_1}}\otimes {\mathbb I}) \circ \Delta^{(M_1)} \circ \tau ,\label{eq:recurPhi}\\
\widehat{B}^{(1)}(u) &=& \sum_{j=1}^{{\mathfrak n}}e^t_{j} \otimes d^{(1)}_{1j}(u)
\end{eqnarray}
where $\bar{\pi}^{(2)}_{a}$ is the fundamental representation evaluation 
homomorphism normalized as in (\ref{repf}), 
$v^{(2)}$ is the application of the highest weight vector 
$e_{2}$ for the space $A_1$:
\begin{eqnarray}
v^{(2)}(X)=X\,(e_{2})^{\otimes 
M_{k-1}}
\end{eqnarray}

The proof is given in \cite{BR2}.

\subsection{Eigenvalues and Bethe Equations}

We state the following commutation relation between $d(u)$ and $\Phi^{{\mathfrak n}}_M(\{u\})$ for $K^+(u)={\mathbb I}$:
\begin{eqnarray}
&&d(u)\Phi^{{\mathfrak n}}_M(\{u\})=U.W.T \nonumber \\
&&+ \Phi^{{\mathfrak n}}_M(\{u\}) \Big(\sum_{k=1}^{{\mathfrak n}}\frac{2u-{\mathfrak n}\hbar}{2u-k\hbar}
\prod_{i=1}^{M_k}f(u-\frac{k\hbar}{2},u_{kj})\prod_{i=1}^{M_{k-1}}\widetilde{f}(u-\frac{(k-1)\hbar}{2},u_{k-1j})
d^{(k)}_{kk}(u-\frac{(k-1)\hbar}{2})\Big)\nonumber \\
&& f(u,v)=\frac{(u-v+\hbar)(u+v+\hbar)}{(u-v)(u+v)}, \quad \widetilde{f}(u,v)=\frac{(u-v-\hbar)(u+v-\hbar)}{(u-v)(u+v)}\label{bigcom}
\end{eqnarray}
with the convention $M_{0}=M_{{\mathfrak n}}=0$.
The $U.W.T$ contains terms with $u$ in the vector. 
We will prove the following theorem:
\begin{theorem}
For $K^+(u)={\mathbb I}$ we have:
\begin{eqnarray}
d(u)\Phi^{{\mathfrak n}}_M(\{u\}) \Omega=\Lambda(u)\Phi^{{\mathfrak n}}_M(\{u\}) \Omega
\end{eqnarray}
If the following set of Bethe equations is satisfed:
\begin{eqnarray}
&&\frac{{\cal K}_{k}(u_{kj}+\frac{k\hbar}{2}) \lambda_{k} (u_{kj}+\frac{k\hbar}{2})  \lambda'_{k} (-u_{kj}-\frac{k\hbar}{2}) }{{\cal K}_{k+1}(u_{kj}+\frac{k\hbar}{2}) \lambda_{k+1} (u_{kj}+\frac{k\hbar}{2})  \lambda'_{k+1} (-u_{kj}-\frac{k\hbar}{2})} = \frac{2u_{kj}}{2u_{kj}+\hbar}\times \nonumber \\
&& \qquad \prod_{i=1}^{M_{k-1}}\frac{(u_{kj}-u_{k-1,i}-\frac{\hbar}{2})(u_{kj}+u_{k-1,i}-\frac{\hbar}{2})}{(u_{kj}-u_{k-1,i}+\frac{\hbar}{2})(u_{kj}+u_{k-1,i}+\frac{\hbar}{2})}
\prod_{i\neq j}^{M_k} 
\frac{(u_{kj}-u_{ki}-\hbar)(u_{kj}+u_{ki}-\hbar)}{(u_{kj}-u_{ki}+\hbar)(u_{kj}+u_{ki}+\hbar)} \times\nonumber \\
&& \qquad\prod_{i=1}^{M_{k+1}} 
\frac{(u_{kj}-u_{k+1,i}-\frac{\hbar}{2})(u_{kj}+u_{k+1,i}-\frac{\hbar}{2})}{(u_{kj}-u_{k+1,i}+\frac{\hbar}{2})(u_{kj}+u_{k+1,i}+\frac{\hbar}{2})},
\nonumber \\
&& j=1,\ldots,M_{k}\,,\quad k=1,\ldots,{\mathfrak n}-1,
\label{BE}
\end{eqnarray}
then the eigenvalues of the transfer matrix have the form:
\begin{eqnarray}
\Lambda(u)&=&\sum_{k=1}^{{\mathfrak n}}\frac{2u-{\mathfrak n}\hbar}{2u-k\hbar}{\cal K}_{k}(u) \lambda_{k} (u)  \lambda'_{k} (-u)\prod_{j=1}^{M_{k}}  
\frac{(u-u_{kj}-\frac{k\hbar}{2}+\hbar)(u+u_{kj}-\frac{k\hbar}{2}+\hbar)}{(u-u_{kj}-\frac{k\hbar}{2})(u+u_{kj}-\frac{k\hbar}{2})}
\nonumber \\
 && \times 
\prod_{j=1}^{M_{k-1}}\frac{(u-u_{k-1j}-\frac{(k-1)\hbar}{2}-\hbar)(u+u_{k-1j}-\frac{(k-1)\hbar}{2}-\hbar)}{(u-u_{k-1j}-\frac{(k-1)\hbar}{2})(u+u_{k-1j}-\frac{(k-1)\hbar}{2})}
\end{eqnarray}
\end{theorem}
\underline{Proof:}\ 
For ${\mathfrak n}=2$ we find the result of section 5.
We will prove the case ${\mathfrak n}+1$ assuming the case ${\mathfrak n}$ is true.
We decompose the transfer matrix:
\begin{eqnarray}
d(u) &=& d_{11}(u)+d^{(2)}(u)\,,\nonumber \\
d^{(2)}(u) &=& tr\Big(D^{(2)}(u)\Big)
\end{eqnarray}
We make a transformation of the operator and a shift of the
spectral parameter to have symmetric commutation relations:
\begin{eqnarray}
d_{11}(u+\frac{\hbar}{2})&=&\widehat d_{11}(u)\,, \qquad
B^{(1)}_a(u+\frac{\hbar}{2})\,=\,\widehat B^{(1)}_a(u), \nonumber \\
D^{(2)}_a(u+\frac{\hbar}{2})&=&\widehat D^{(2)}_a(u)-
\frac{\hbar}{2u} \,{\mathbb I}^{(2)}_a  \otimes \widehat d_{11}(u). 
\end{eqnarray}
From this transformation we get a new form for the transfer matrix:
\begin{eqnarray}
d(u+\frac{\hbar}{2}) &=&\frac{2u-{\mathfrak n}\hbar}{2u}\widehat d_{11}(u)+tr_a\Big(\widehat D_a^{(2)}(u)\Big),
 \end{eqnarray}
The commutation relations  
between $\widehat d_{11}(u)$, $\widehat D^{(2)}(u)$ and 
$ \widehat B^{(1)}(u)$ are obtains from reflection equation (\ref{ER}):
  \begin{eqnarray}
\widehat B^{(1)}_a(u)\,\widehat B^{(1)}_b(v)&=& 
\widehat B^{(1)}_b(v)\, \widehat B^{(1)}_a(u)\,\mbox{${\mathbb R}$}_{ab}^{(2)}(u-v), 
\label{eq:comBB}\\
\widehat d_{11}(u)\, \widehat B^{(1)}_b(v)&=&
\frac{(u-v+\hbar)(u+v+\hbar)}{(u-v)(u+v)}
\widehat B_b(v) \widehat d_{11}(u) -\frac{\hbar(2v+\hbar)}{(u-v)2v}\widehat B^{(1)}_b(u) 
\,\widehat d_{11}(v) 
\nonumber \\
&& +\frac{\hbar}{u+v}\, \widehat B^{(1)}_b(v) \widehat D^{(2)}_b(v), 
\label{eq:comdkkB} \\
 \widehat D^{(2)}_a(u)\widehat B^{(1)}_b(v)&=&\frac{(u-v-\hbar)(u+v-\hbar)}{(u-v)(u+v)}\widehat B^{(1)}_b(v) 
 \mbox{${\mathbb R}$}^{(2)}_{ab}(u+v)
 \widehat D^{(2)}_a(u) \mbox{${\mathbb R}$}^{(2)}_{ab}(u-v)\nonumber \\
&&-\frac{\hbar(2v+\hbar)}{4uv(u+v)}\widehat 
B^{(1)}_b(u)R^{(2)}_{ab}(2u)P^{(2)}_{ab}\,\widehat d_{kk}(v)\nonumber \\
&&+\frac{\hbar}{(u-v)2u}\widehat B^{(k)}_b(u)
R^{(2)}_{ab}(2u)
\widehat D^{(2)}_a(v)P^{(2)}_{ba}.
\label{eq:comDB}
\end{eqnarray}

From this commutation relations and using the fact that the Bethe vector is globally invariant
if we permute $\widehat{B}$ we obtain two types of terms: the wanted and unwanted.
Let us consider first the wanted terms.
For $\widehat d_{11}(u)$ we have:
\begin{eqnarray}
\prod_{i=1}^{M_1}f(u,u_{1i})\Phi^{{\mathfrak n}+1}_{M}(\{u\})\widehat d_{11}(u)\Omega
\end{eqnarray} 
where we have used the theorem \ref{theo:embedding} to put $\widehat d_{11}(u)$ in the right.
For $\widehat D^{(2)}(u)$ we have:
\begin{eqnarray}
\prod_{i=1}^{M_1}\widetilde{f}(u,u_{1i})\widehat B^{(1)}_{a^1_1}(u_{11}) \cdots 
 \widehat  B^{(1)}_{a^1_{M_1}}(u_{1M_1})\,\widehat\Psi^{(1)}_{\{u_1\}}
\Big(d(u)\Phi^{{\mathfrak n}}_{M-M_1}(\{u\}/\{u_1\})\Big),
\end{eqnarray}
where $d(u)$ is the transfer matrix for ${\mathfrak D}_{{\mathfrak n}}$. We have used the definition of $\widehat\Psi^{(1)}_{\{u_1\}}$ and the relations (\ref{tau},\ref{repf}) to find:
\begin{eqnarray}
\prod_{i=1...M_1}^{\longrightarrow} \mbox{${\mathbb R}$}^{(2)}_{aa_i^1}(u+u_{1i})
 \widehat D^{(2)}_a(u) \prod_{i=1...M_1}^{\longleftarrow}\mbox{${\mathbb R}$}^{(2)}_{aa_i^1}(u-u_{1i})\widehat\Psi^{(1)}_{\{u_1\}}(X)=\widehat\Psi^{(1)}_{\{u_1\}}(D_a(u)X)
\end{eqnarray}
Using (\ref{bigcom}) we can commute $d^{(2)}(u)$. It remains to compute the action of  $\widehat\Psi^{(1)}_{\{u_1\}}$.
The formulas (\ref{tau-bis},\ref{repf}) and the fact that $\widehat\Psi^{(1)}_{\{u_1\}}$ is an morphism (up to $\nu^{(2)}$) allow to find ($1<k\leq{\mathfrak n}$):
\begin{eqnarray}
\widehat\Psi^{(1)}_{\{u_1\}}(d^{(1)}_{11}(u))&=&d^{(2)}_{22}(u)\nonumber \\
\widehat\Psi^{(1)}_{\{u_1\}}(d^{(k)}_{kk}(u-\frac{(k-1)\hbar}{2}))&=&d^{(k+1)}_{k+1k+1}(u-\frac{(k-1)\hbar}{2})\prod_{i=1}^{M_1}\widetilde{f}^{-1}(u,u_{1i})  \label{phi-act}
\end{eqnarray}
Using these formulas and making a reverse shift, the theorem is proved for the wanted term.

Let us now consider the unwanted terms. 
First we give the reason why it is not possible to deal with a general diagonal $K^+(u)$ matrix.
To factorised the unwanted term for $\widehat d_{11}(u)$ we must have $tr((K_a^+)^{(2)}(u)R^{(2)}_{ab}(2u)P^{(2)}_{ab})\propto {\mathbb I}^{(2)}$. The only possibility is $(K_a^+)^{(2)}(u)\propto {\mathbb I}^{(2)}$.
Here we will just prove the case $K^+(u)= {\mathbb I}$ and let the reader consult \cite{BR2}
 for the other case: $K^+(u)= k(u)E_{11}+{\mathbb I}^{(2)}$.
 
 Using the commutation relations and looking to the term with $\widehat d_{11}(u_{11})$--the other term
 are similar using the invariance by permutation of the Bethe vector, see theorem \ref{trace}--we find: 
\begin{eqnarray}
-\frac{(2u-{\mathfrak n}\hbar)(2u_{11}+\hbar)}{2u_{11}(u^2-u_{11}^2)}\prod_{i=2}^{M_1}f(u_{11},u_{1i})\Phi^{{\mathfrak n}+1}_{M}(\{u\},u_{11}\to u)\widehat d_{11}(u_{11}) \nonumber
\end{eqnarray} 
Looking now for the term $\widehat D^{(2)}(u_{11})$, after using the trick  $tr(R^{(2)}_{ab}(2u)P^{(2)}_{ab})=(2u-{\mathfrak n}\hbar){\mathbb I}^{(2)}$ to obtain a good form for commuting with $\widehat\Psi^{(1)}_{\{u_1\}}$, we find:

\begin{eqnarray}
\frac{(2u-{\mathfrak n}\hbar)(2u_{11}-\hbar)\hbar}{(2u_{11}-{\mathfrak n}\hbar)(u^2-u_{11}^{2})}\prod_{i=2}^{M_1}\widetilde{f}(u_{11},u_{1i})\widehat B^{(1)}_{a^1_1}(u) \cdots 
 \widehat  B^{(1)}_{a^1_{M_1}}(u_{1M_1})\,\widehat\Psi^{(1)}_{\{u_1\}}
\Big(d(u_{11})\Phi^{{\mathfrak n}}_{M-M_1}(\{u\}/\{u_1\})\Big)\nonumber
\end{eqnarray}

From (\ref{phi-act}) we see that only the first term of the eigenvalue is non zero. We can obtain
the Bethe equation for $u_{11}$:
\begin{eqnarray}
\frac{{\cal K}_1(u_{11}+\frac{\hbar}{2})\lambda_1(u_{11}+\frac{\hbar}{2})\lambda'_{1}(u_{11}+\frac{\hbar}{2})}{{\cal K}_2(u_{11}+\frac{\hbar}{2})\lambda_2(u_{11}+\frac{\hbar}{2})\lambda'_{2}(u_{11}+\frac{\hbar}{2})}=\frac{2u_{11}}{(2u_{11}+\hbar)}
\prod_{i=2}^{M_1}\frac{\widetilde{f}(u_{11},u_{1i})}{f(u_{11},u_{1i})}\prod_{i=1}^{M_2}f^{-1}(u_{11}-\frac{\hbar}{2},u_{2i})\nonumber 
 \end{eqnarray}
 Using the invariance by permutation the Bethe equations for the other $u_{ij}$ follow.
 We must also modify the other Bethe equations. The only change comes from the relations (\ref{phi-act}) who change the eigenvalues of the 
$d^{(k)}_{kk}(u-\frac{(k-1)\hbar}{2})$. This modification only affects the first familly of Bethe equations adding a term to the right product.
This ends the recursion and proves the theorem.
\null \hfill {\rule{5pt}{5pt}}\\[2.1ex]\indent

\newpage

\section{Conclusion}
In this proceeding we give the Nested Bethe ansatz for open spin chains of XXX-type with diagonal boundary conditions.
This result could be extend to the case of non diagonal boundary conditions but with some constraints  between
$K^+(u)$ and $K^-(u)$. To do this, we use the $GL({\mathfrak n},{\mathbb C})$ invariance of the Yangian \cite{GM,AACDFR} and take for an arbitrary invertible
$M$:
\begin{eqnarray}
\widetilde K^+(u)=MK^+(u)M^{-1}, \quad \widetilde K^-(u)=MK^-(u)M^{-1}
\end{eqnarray}
It's equivalent to the assertion that $K^+(u)$ and $K^-(u)$ are diagonalisable in the same basis, otherwise the nested Bethe ansatz does not work and the diagonalisation of the transfer matrix remains an open problem.
 
We also give a \textbf{trace formula} for the Bethe vector of the 
open chain. This formulation could be a starting point for the investigation 
of the quantized Knizhnik-Zamolodchikov equation following the work \cite{TV2}.
For such a purpose, the coproduct properties of Bethe vectors for open 
spin chains remain to be studied.
Defining a scalar product and computing the norm of these Bethe 
vectors is also a point of  fundamental interest.

 \section*{Acknowledgments:} The first autor thanks E. Ivanov and S. Fedoruk for invitation at the Dubna International SQS'09 Workshop ("Supersymmetries and Quantum Symmetries-2009", july 29 - August 3, 2009) where this work has been presented and F.Ravanini for comments on the manuscript.
This work was supported by the INFN Iniziativa Specifica FI11.

\end{document}